\documentstyle[12pt,aasms4,flushrt]{article}
\newcommand{\mincir}{\ \raise -2.truept\hbox{\rlap{\hbox{$\sim$}}\raise5.truept
	\hbox{$<$}\ }}			
\newcommand{\magcir}{\ \raise -2.truept\hbox{\rlap{\hbox{$\sim$}}\raise5.truept
 	\hbox{$>$}\ }}		
\lefthead{Giallongo et al.}
\righthead{The Proximity effect from a large Lyman-$\alpha$ sample}

\begin{document}
\title{ THE PROXIMITY EFFECT, THE UV BACKGROUND AND THE STATISTICS
OF THE LYMAN-ALPHA LINES AT HIGH RESOLUTION \footnote{Based on
observations collected at the 3.5m New Technology Telescope of ESO for
the Key program 2-013-49K}}

\bigskip
\bigskip

\author{E. Giallongo$^2$, S. Cristiani$^{3,4}$, S. D'Odorico$^4$, 
A. Fontana$^2$, S. Savaglio$^{4,5}$}

~

\bigskip
\bigskip
\bigskip
\bigskip

\noindent
$^2$ Osservatorio Astronomico di Roma, via dell'Osservatorio, I-00040
Monteporzio, Italy\\
\noindent
$^3$ Dipartimento di Astronomia, Universit\`a di Padova, vicolo
dell'Osservatorio 5, I-35122, Padova, Italy\\
\noindent
$^4$ European Southern Observatory, Karl Schwarzschild Strasse 2,
D-85748 Garching, Germany\\
\noindent
$^5$ Istituto di Astrofisica Spaziale del CNR, Frascati, Italy

\newpage
\begin{abstract}
We present results from high resolution (R$\simeq 28000$) spectra of
six high-redshift QSOs taken at the ESO NTT telescope that allow the
detailed study of the Lyman-$\alpha$ population in the redshift
interval $z=2.8-4.1$.

The typical Doppler parameters found for the Lyman-$\alpha$ lines lie
in the interval $b = 20 \div 30$ km s$^{-1}$, corresponding to
temperatures $T > 24000$ K, with a fraction of the order 15\% in the
range $ 10 \le b \le 20$ km s$^{-1}$.  These values are still
consistent with models of low density, highly ionized clouds.

The observed redshift and column density distributions obtained from
these spectra and from the observations of 4 additional QSOs taken in
the literature allow an accurate estimate of the proximity effect from
a relatively large Lyman-$\alpha$ sample (more than 1100 lines with
$\log N_{HI}\geq 13.3$) in the redshift interval $z=1.7-4.1$.

A Maximum Likelihood analysis has been applied to estimate {\it
simultaneously} the best fit parameters of the Lyman-$\alpha$
statistics {\it and} of the UV background.  After correcting for the
blanketing of weak lines, we confirm that the column density
distribution is best represented by a double power-law with a break at
$\log N_{HI}\simeq 14$, with a slope $\beta_s = 1.8$ for higher column
densities and a flatter slope $\beta_f = 1.4$ below the break.

A value $J_{LL} = 5\pm 1 \times 10^{-22}$ erg cm$^{-2}$ s$^{-1}$
Hz$^{-1}$ sr$^{-1}$ is derived for the UV background in the redshift
interval $z=1.7-4.1$, consistent with the predicted QSO
contribution. No evidence is found for redshift evolution of the UVB
in the same redshift interval.

The comoving volume density distributions of protogalactic damped
systems, Lyman Limit systems and Lyman-$\alpha$ clouds with $\log
N_{HI}\magcir 14$ and radii $R\simeq 200$ kpc are found to be similar,
suggesting a possible common association with galaxies.
\end{abstract}

\section{Introduction}
The statistics of the Lyman-$\alpha$ absorptions, seen in QSO spectra
shortward of the Lyman-$\alpha$ emission, provide unique information
about the physical and cosmological properties of the neutral gas
phase of the baryonic component of the Universe.  The high sensitivity
of the observations of the Lyman-$\alpha$ lines allows to investigate
very different structures, ranging from fluctuations of the diffuse
intergalactic medium to the interstellar medium in protogalactic
disks.  This scientific driver coupled with the advances in
instrumental and detector capabilities at 4-m class telescope has led
in the last few years to a flourishing of investigations on
Lyman-$\alpha$ forest (e.g. Pettini et al. 1990, Carswell et al. 1991,
Rauch et al. 1993, Giallongo et al. 1993, Fan \& Tytler 1994,
Cristiani et al. 1995). The new data have sufficient resolution
(R$>$20000) and signal-to-noise ratio to use line-fitting procedures
to obtain reliable estimate of the L-$\alpha$ clouds parameters like
redshift, column density $N_{HI}$ and Doppler width $b$ and a more
detailed physical picture of the absorbing gas has emerged.

The average Doppler parameter is found to be $b\mincir 30$ km
s$^{-1}$, corresponding to an average temperature $T\sim 5\times 10^4$
K. This is typical for clouds in photoionization equilibrium with the
general UV ionizing background.  However, the presence of a non
negligible fraction of narrow lines with $b=10-20$ km s$^{-1}$ has
been recognized in the highest resolution spectra, although the
assessment of the real fraction relies on the S/N ratio of the data
and on the intrinsic blending of the lines, which is strong at very
high redshifts. It is difficult to reconcile, using standard
photoionization models, temperatures $T\sim 15000-20000$ K with the
large cloud sizes $R\sim 50-150h^{-1}$ kpc inferred from observations
of QSO pairs (Smette et al. 1992, 1995; Bechtold et al.  1994; Dinshaw
et al. 1995). This has led to the introduction of particular cooling
mechanisms (Giallongo \& Petitjean 1994).

The Lyman-$\alpha$ lines show an appreciable redshift evolution in the
redshift range $z=1.7-4$ and small clustering, limited at $\Delta
v<300$ km s$^{-1}$ for lines with $\log N_{HI}\magcir 13.8$ (Cristiani
et al. 1995).  Attempts have been done to explain these cosmological
properties in the framework of the CDM cosmology
(e.g. Miralda-Escud\'e \& Rees 1994, Cen et al. 1994).

The introduction of the HIRES echelle spectrograph at the Keck 10m
telescope has given new impulse to the observations.  Recent spectra
with HIRES have revealed the presence of CIV absorption associated with
Lyman-$\alpha$ lines with $\log N_{HI} \magcir 14$ (Cowie et al. 1995,
Tytler \& Fan 1995). The derived abundances, although dependent on the
ionization state of the Lyman-$\alpha$ clouds, seem to be similar to
the ones derived for the heavy element absorptions originated in
galaxy halos, suggesting a continuity in their physical properties.

The high resolution observations of the Lyman-$\alpha$ forest provide
also a powerful method to estimate the UV background at high
redshifts.  It is already known from low resolution Lyman-$\alpha$
samples that the redshift evolution of the line number density, within
a single spectrum, does not follow the general cosmological trend when
approaching the QSO emission redshift. This ``inverse effect'' has
been interpreted and modeled as a ``proximity effect'' (Weymann,
Carswell, \& Smith 1981; Tytler 1987; Bajtlik, Duncan \& Ostriker
1988).  It consists in a reduction of the line density in the region
near the QSO emission redshift due to the increase of the ionizing
flux by the QSO.  In this way, absorbers near QSOs are more highly
ionized than those farther away where the general UV background is the
only source of ionization.

Using a simple photoionization model Bajtlik, Duncan, \& Ostriker 1988
showed that the predicted and observed proximity effects agreed at
high redshift adopting the value $J_{LL}\sim 10^{-21}$ erg cm$^{-2}$
s$^{-1}$ Hz$^{-1}$ sr$^{-1}$ for the UV background at the Lyman
limit. This estimate has been confirmed within uncertainties of $\pm
0.5$ in $\log J_{LL}$ by Lu, Wolfe \& Turnshek 1991. The most
comprehensive analysis of the proximity effect has been published by
Bechtold 1994 on the basis of a homogeneous low resolution (R$\sim
$3500) Lyman-$\alpha$ sample derived from the spectra of 34 QSOs in
the redshift range $1.6<z<4.1$. She found a value about 3 times
higher, $J_{-21}=3$ and a correlation of the proximity effect on
the QSO luminosities as expected from the photoionization
model. However, any analysis performed at low resolution does not
allow the use of the correct column density and redshift distributions
for the Lyman-$\alpha$ lines, both affecting the estimate of the UV
background from the proximity effect.  Giallongo et al. (1993) and
Cristiani et al. (1995) on the basis of a few quasars observed at high
resolution (R$\magcir 25000$), suggest a rather flat power-law column
density distribution for lines with $\log N_{HI}<14$ and a value of
the UV background sensibly smaller than the reference value $\log
J_{LL}=-21$ at $z\sim 3$, even if the uncertainties were relatively
large.

We discuss in this paper a larger sample including more than 1100
Lyman-$\alpha$ lines with column densities $\log N_{HI}\geq 13.3$,
observed at an average R$\sim 25000$, and perform a Maximum Likelihood
analysis to estimate {\it simultaneously} the parameters of the
Lyman-$\alpha$ line distributions {\it and} the intensity of the UV
background in the redshift interval $1.7<z<4.1$. In particular, in
section 2.1 our data sample is defined, in section 2.2 the $b-\log
N_{HI}<14$ distribution and its implication on the temperature of the
Lyman-$\alpha$ clouds are analysed. In section 2.3 and 2.4 we discuss
the H I column density and redshift distribution respectively. Section
3 gives the details of the computation and the results on the UV
background. The conclusions of our work are listed in section 4.

We adopt throughout the value H$_o = 50$ km s$^{-1}$
Mpc$^{-1}$ for the Hubble parameter and $q_o = 0.5$.

\section{Lyman-$\alpha$ Statistics}
\subsection{The Data Sample}

Most of the data used in this analysis come from the ESO key programme
2-013-49K (P.I. S.D'Odorico) devoted to the study of the
intergalactic medium at high redshifts.  Up to now spectra of six
QSOs, at a resolution between 9 and 14 km s$^{-1}$, have been reduced.
The list of the objects, with emission redshifts ranging from 3.3 to
4.1 is given in Table 1 together with the main characteristics of the
Lyman-$\alpha$ absorption line sample.  Spectra of the QSOs 2126-158
and 0055-269 have been already published (Giallongo et al. 1993;
Cristiani et al. 1995).  For 2126-158 new spectra have been added,
significantly improving the S/N with respect to the published data.  A
complete description of the data sample will be given in a separate
paper: here we focus on the main statistical properties of the
Lyman-$\alpha$ clouds that can be derived from this sample.

All the spectra have been analyzed in a uniform way and all the lines
have been fitted using our FITLYMAN program which is now available in
the ESO-MIDAS package (Fontana \& Ballester 1995).  It performs a
$\chi^2$ minimization to derive the redshift $z$, the Doppler
parameter $b$ and the column density $N$ for isolated lines and
individual components of the blends.  As done in previous similar
analyses (e.g. Giallongo et al. 1993), complex structures with often
asymmetric profiles have been fitted with the minimum number of
components required to give a probability of random deviation
$P>0.05$.

To identify heavy element systems we have compared the lists of the
observed lines with a list, derived from Morton 1991, containing the
most frequently seen lines in QSO absorption spectra.  As customary,
we searched for significant excesses of identifications at all
possible redshifts (Bahcall 1968).  All the identified heavy element
systems present in the Lyman-$\alpha$ forest have been removed from
the final sample, which consists of more than 1100 Lyman-$\alpha$
lines with observed column density $\log N_{HI}\geq 13.3$.  Whenever
possible, the Lyman-$\beta$ forest has been used to constrain the
number of components in the strong saturated Lyman-$\alpha$ blends as
shown in Fig. 1.

The S/N has been computed from the noise spectrum, and is typically
greater than 10 per pixel element (corresponding to 15 per resolution
element).  In the regions near the QSO Lyman-$\alpha$ emissions, where
the proximity effect is important, and for the brighter quasars the
S/N raises to higher values $\sim 15-40$, corresponding to S/N$\sim
20-60$ per resolution element.  Extensive simulations by various
authors (Rauch et al. 1993, Fontana \& Ballester 1995) show that no
strong bias in the derived statistical distributions is expected when
the S/N is $>$10 and $\log N_{HI}>13$.

We have complemented our data with other spectra available in the
literature with similar resolution and redshift range obtaining a
final sample of 10 QSOs (in the following referred to as {\it extended
sample}), that is $\sim 30$\% of the number of objects used by
Bechtold (1994) , but observed at resolution ten times higher.

\subsection {The $b-N_{HI}$ Distributions and the 
Temperature of the Lyman-$\alpha$ Clouds}

The average resolution ($R\sim 25000$) of our extended sample
corresponds, in terms of Voigt profile fitting, to a minimum Doppler
parameter $b\simeq 7$ km s$^{-1}$, which is equivalent to a
temperature $T\sim 3000$ K and thus adequate to assess the
photoionization state of the Lyman-$\alpha$ clouds.

If we restrict our analysis to the two objects with the best S/N ratio
(20-60 per resolution element), 0000-26 and 2126-158 (Fig.2), we can
obtain reliable measurements of the Doppler profiles of the weak lines
down to $\log N_{HI} \mincir 13$.  In Figs. 3 and 4 we show the
$b-N_{HI}$ distributions derived from 0000-26 and 2126-158
respectively, while Fig. 5 gives the histograms of the Doppler
parameter.

These plots confirm the earlier result (Giallongo et al. 1993,
Cristiani et al. 1995) that the bulk of the line distribution lies in
the $b$ region between 20 and 30 km s$^{-1}$.  There is a
non-negligible fraction of lines with $b$ in the range 10--20 km
s$^{-1}$, 18\% and 17\% for 0000-26 and 2126-158 respectively, with
very few cases ($\mincir 2$\%) of lines with $b<10$ km s$^{-1}$, which
may be representative of unrecognized metal lines. Since the profile
fitting of narrow lines with $\log N_{HI}<13$ can be affected by
systematic errors due to the present S/N (Rauch et al. 1993) we have
computed the same fraction for lines with $\log N_{HI}\geq 13$ finding
16\% and 15\% in both spectra.  In any case, the fraction of low-b
lines appears to be uncorrelated with the S/N.

It is well known (e.g. Pettini et al. 1990, Donahue \& Shull 1991)
that Doppler $b$ values corresponding to temperatures $T\mincir 20000$
K are difficult to reconcile with sizes greater than 50 kpc.
Giallongo \& Petitjean 1994 emphasized the importance of inverse
Compton cooling by the cosmic microwave background radiation and of a
steepening of the ionizing background at wavelengths shorter than the
HeII edge to get sufficient cooling rate to obtain temperatures as low
as $18000-20000$ K with sizes up to $100-200$ kpc.  In all these
analyses both photoionization and thermal equilibrium were assumed,
neglecting explicit time dependence of the photoionization mechanisms,
which may be important at redshifts $z\magcir 3$.

Ferrara \& Giallongo 1995 recently developed a time dependent
photoionization code that follows the thermal history of the
photoionized gas with total density $n_{_{H}}\mincir 10^{-4}$
cm$^{-3}$ out of the ionization and thermal equilibrium.  They show
that it is possible to get temperatures as low as 15000 K at $z>$3
provided that the ionization starts at redshifts appreciably higher
than 7 and that the UVB has a jump at the HeII edge by a factor of the
order of 100. This corresponds to minimum Doppler values of the order
of $b\sim 14$ km s$^{-1}$, consistent with the minimum values observed
in our sample.  Thus, the existence of narrow lines in the
Lyman-$\alpha$ population may become a sensitive tool to constrain the
epoch of reionization and the shape of the UVB.

\subsection{The HI column density distribution}

The column density distribution of the lines in our extended sample at
a distance larger than 8 Mpc from the QSOs is shown in Fig. 6, in
which the line density has been normalized at $z=3$.

A double power-law distribution clearly appears with a break at $\log
N_{HI} = 14$.  Below the break the slope is very flat ($\beta_f \sim
1.1$) down to our selection threshold $\log N_{HI}\simeq 13$.  Above
the break the slope is steep with $\beta_s \sim 1.8$.  There are two
main biases that do alter the shape of the column density
distribution. Strong saturated lines above the break often show
complex structures, but independently of the S/N, these features can
not be deblended without information on the Lyman-$\beta$ forest,
which are often outside the observed range or heavily contaminated by
lower-$z$ Lyman-$\alpha$ lines.

The distribution of the lines below the break is affected by the
so-called ``line-blanketing'' effect, due to high column density lines
that conceal weak lines.  In order to quantify and correct this
effect, which becomes more and more important at higher redshifts, we
have performed extensive simulations with the FITLYMAN code, trying to
mimic the blanketing process.  Each set of simulations was performed
by progressively merging two lines (a ``weak'' and a ``strong'' line)
of given parameters: fig. 7a shows an example at S/N=10. At each
separation, a profile fit with a single component was attempted.  We
have defined the critical separation
as the distance for which the fit turns out to be acceptable in the
50\% of trials (about 0.7 {\AA} in this case, see fig. 7b).  At
smaller separations, with the assigned S/N ratio, resolution and
redshift, the weak line is definitely lost.
In Figs. 7c,d a similar simulation is shown for a line pair with
larger $b$ parameters. As expected, in this case the critical
separation is larger, $\sim 1$ {\AA}.
This critical separation, which is a function of several parameters
(the S/N ratio, the Doppler parameters of the lines and their column
densities, the average absorption redshift and the resolution) was
computed for a grid in the parameter space corresponding to the
available data. In this way we have been able to quantify the effects
of the blanketing on the line selection function.
Lines with $\log N_{HI}=13.3-13.5$ are lost in the $\sim$35\% of the
observed redshift interval at $z\simeq 4$, in the $\sim$17\% at
$z\simeq 3$ and in the $\sim$11\% at $z\simeq 2$.

The effects of this correction on the column density distribution are
shown in fig. 6 (dashed histogram), where the number density of the
lines in the different column density bins has been normalized to the
effective redshift range derived for each QSO spectrum. The change in
slope is still present and well represented by a double power-law,
although the average slope below the break is increased to $\beta_f
\sim 1.4$, a value consistent with previous estimates obtained at
lower redshifts (Giallongo et al. 1993, Cristiani et al. 1995) where
line blanketing effects are less important.  The corrected
distribution has been computed down to $\log N_{HI}=13.3$ which is the
completeness threshold of our extended sample.

\subsection{The Redshift Distribution}

The observed number density of the extended sample of Lyman-$\alpha$
clouds with $\log N_{HI}\geq 14$ is plotted in Fig. 8 as a function of
redshift together with that of the Lyman Limit and Damped systems
recently derived by Storrie-Lombardi et al. 1994; 1995. The number
density of Ly$\alpha$ clouds at $z<1.7$ derived from the HST data of
Bahcall et al. 1995 is also plotted.

Assuming a standard power-law evolution of the type $dn/dz\propto
(1+z)^{\gamma}$, Storrie-Lombardi et al. found $\gamma=1.5$ for 
both populations, while we find (see Section 3) $\gamma=2.7$ for the
Lyman-$\alpha$ clouds. Of course, the derived number densities span
about two orders of magnitude.

Estimates of the comoving volume densities can be derived using
information about the sizes of the different populations.  Bechtold et
al. 1994 give, assuming spherical geometry, a median value for the
radius of Lyman-$\alpha$ clouds of the order of 200 kpc. The Lyman
Limit population is virtually indistinguishable from the population
selected by the presence of MgII absorption which is found to be
associated with the internal parts of the galaxy halos (Bergeron \&
Boiss\'e 1991). From the statistics of the galaxies responsible for the
MgII absorptions it is possible to derive an average radius of the
order of 40 kpc (Steidel 1995).

For the damped systems, typical HI column densities $\geq 2 \times
10^{20}$ cm$^{-2}$ imply disks of $ \sim 20 $ kpc radii (Smette et
al. 1995; Broeils \& van Woerden 1994; Steidel et al. 1995).  With
such sizes and geometries the volume densities of the Lyman-$\alpha$
clouds with $\log N_{HI}\magcir 14$, i.e. at the break of the column
density distribution, and of the Lyman Limit and Damped systems appear
remarkably similar, as shown in Fig. 9.  Also their cosmological
evolutions appear consistent in the overall redshift range.

Thus, it is tempting to interpret the Lyman-$\alpha$ clouds with $\log
N_{HI} \magcir 14$ as gas associated with the outer regions of the same
class of galaxies responsible for the Lyman Limit and Damped systems.

A Lyman-$\alpha$-galaxy association was originally proposed by Bahcall
\& Spitzer 1969 and it is now statistically suggested at low redshift
by imaging of the galaxy-fields surrounding QSOs (Lanzetta et
al. 1995).

A scenario of this kind is in agreement with the observations of
metallicities of the order of $10^{-2}$ in the Lyman-$\alpha$ clouds
(Cowie et al. 1995, Tytler \& Fan 1995) and also with the detection of
clustering of the Lyman-$\alpha$ clouds with $\log N_{HI}\magcir 14$
(Cristiani et al. 1995; Meiksin \& Bouchet 1995).

The Lyman-$\alpha$ absorptions with $\log N_{HI} \mincir 14$ (the flat
part of the column density distribution), would be in part associated
with galaxies and in part due to fluctuations of a clumpy intergalactic
medium.

\section{The measure of the UV Background from the proximity effect}

We can parameterize the line distribution far away from the QSOs as a
function of the column density and redshift:

\begin{equation}
{\partial ^2 n \over \partial z\partial N_{HI}}= A_o (1+z)^{\gamma} 
\left\{ \begin{array}{ll}
N_{HI}^{-\beta _f} & N_{HI} < N_{break}\\
N_{HI}^{-\beta _s} N_{break}^{\beta _s - \beta _f} 
& N_{HI} \geq N_{break} \end{array} \right.
\end{equation}

Near the QSO, highly ionized clouds are observed with a column density 
\begin{equation}
N_{HI}={N_{\infty} \over {1+\omega}}
\end{equation}
i.e. the ratio between the intrinsic column density $N_{\infty}$,
which the same cloud would have at infinite distance from the QSO, and
the factor $1+\omega$, where:
\begin{equation}
\omega (z)= {F \over 4 \pi J}
\end{equation}
is the ratio between the flux $F$ that the cloud receives from the QSO
and the flux $J$ that the cloud receives from the general UVB (see
Bajtlik et al. 1988 and Bechtold 1994 for details of the model).

Adopting the following conservation law near the QSO emission redshift we have
\begin{equation}
f(N) = g(N_{\infty}) dN_{\infty}/dN=g(N_{\infty}) (1+\omega)
\end{equation}
where $f(N)$ and $g(N_{\infty})$ are the column density distributions near
the QSO and at infinite distance.

We have used the following double power-law distribution to get a
simultaneous estimate of the Lyman-$\alpha$ parameters of the
$N_{HI}$, $z$ distributions {\it and} of the UV background $J_{LL}(z)$
evaluated at the Lyman limit.
\begin{equation}
{\partial ^2 n \over \partial z\partial N_{HI}}= A_o (1+z)^{\gamma} 
(1+\omega)^{1-\beta _f}
\left\{ \begin{array}{ll}
N_{HI}^{-\beta _f} & N_{HI} < N_{break}\\
N_{HI}^{-\beta _s} N_{break}^{\beta _s - \beta _f} 
& N_{HI} \geq N_{break} \end{array} \right.
\end{equation}
where 
$$N_{break} (z)= {N_{\infty,b} \over 1+\omega(z)}$$ 
is the observed break, which is shifted to lower and lower $N_{HI}$ as
the QSO emission redshift is approached,

It is clear from the previous equation that the measure of $J$ depends
critically on the estimates of $\beta_f$, because the bulk of the
lines comes from this part of the distribution, and on $\gamma$. For
this reason, it is not strictly correct to separate the analysis of
the Lyman-$\alpha$ line distribution from the estimate of $J$, as done
in previous analyses performed at lower resolution, and a Maximum
Likelihood Analysis which adopts the line distribution given in Eq. 5
has to be used.

A similar approach has been adopted by Kulkarni and Fall 1993 using
low redshifts HST spectra.  However, their data are at low resolution
and they derived column densities from a curve of growth analysis
assuming an average Doppler parameter for all the lines and a single
power-law $N_{HI}$ distribution.

In Table 2 we have reported the result of the ML analysis.  The
redshift evolution is confirmed at about the same level previously
found at lower resolution (e.g Lu, Wolfe \& Turnshek 1991, Bechtold
1994). The column density distribution is well represented by a double
power-law, as already shown in Fig. 6, with a very flat slope below
the break at $\log N_{HI}\sim 14$. Another important result concerns
the UV background derived from the proximity effect that appears
definitely smaller than previously reported, i.e. $J_{-22}=6 \pm 1$ in
the redshift interval $z=1.7-4.1$.

The redshifts adopted in Table 1 are derived from low ionization lines
(OI, MgII, H$\alpha$), which are representative of the true systemic
redshift of the QSO (Gaskell 1982, Espey et al. 1989).  Using high
ionization lines (e.g. CIV, SiIV etc.), which typically provide a
smaller redshift, would result in an overestimate of the value of the
UV background derived from the proximity effect by a factor of $\sim
2$ (Espey 1993, Bechtold 1994).

As apparent in equation 5, the value of $J$ is sensitive to the
precise value of $\beta _f$. Thus, it is important to apply a
blanketing correction below the break of the column density
distribution.  To this end, we have introduced for each spectrum in
the Maximum Likelihood Analysis an effective redshift range which is a
function of column density and redshift, as defined in the section
2.3.

The result is shown in Table 3: the correction for the blanketing
effects increases the rate of the redshift evolution and steepens the
column density distribution below the break.  Because of these
changes, the intrinsic number density of lines far away from the QSOs
increases especially at high redshifts, and this in turn causes a
decrement of the value of the UV background derived from the proximity
effect down to $J_{-22}=5\pm 1$.

We have repeated the ML analysis allowing for a power-law redshift
evolution of the UVB of the kind
\begin{equation}
J=J_{(z=3)} \left({1+z\over 4}\right) ^j
\end{equation}
The best fit slope $j$ shown in Table 3 is negative but consistent
with no evolution in the redshift interval $z=2-4$, although the
uncertainties are still large.

The present statistical analysis, based on the ML formalism, allows an
accurate measure of the UV background, even if systematic biases could
be present. Bechtold 1994 gave a comprehensive discussion of the
various systematic errors possibly affecting any estimate of the UVB.
One of the main source of errors concerns the assumption of the same
spectral shape for the UVB and the spectrum of the individual QSOs,
which is implicit in the Bajtlik et al. 1988 model.  Bechtold 1994,
comparing the ionization rate due to an UVB produced by QSOs alone
with the one produced by the sum of QSOs and primeval galaxies,
estimated small changes ($\mincir 25$\%) in the ratio of the
ionization rates.

Recent measurements of HeII Lyman-$\alpha$ absorption in QSO spectra
(Jakobsen et al. 1994, Davidsen et al. 1995) show that a strong
decrease, i.e. a ``jump'' by a factor $\sim 70 -100$, in the shape of
the UVB should be present just above the HeII edge (4 Rydberg).
However, detailed Lyman-$\alpha$ photoionization models by Ferrara \&
Giallongo 1995 show that, even in this case, no appreciable changes in
the hydrogen ionization fractions are present either including or
excluding the jump in the spectral shape of the UV background. Thus
the assumption of the simple power-law shape adopted in this paper and
in Bechtold 1994 remains a valid approximation.

Other biases such as magnification by gravitational lensing could be
present for some QSOs, but the statistical effect is expected to be be
small (Bechtold 1994), on the ground of the correlation of the
proximity effect with luminosity and not with redshift.

QSO variability on a time scale of the order of the photoionization
time ($\sim 10^4$ yr) could also affect the estimate of the proximity
effect, but the result should simply be an increase in the statistical
variance of our measure without changing the average value.

The value of $J$ so derived is a factor of 6 lower than previous
estimates (e.g. Bechtold 1994) from data at low resolution using the
same simple model.  This difference can be ascribed essentially to the
assumption of a single power-law distribution with the canonical slope
$-1.7$ adopted by Bechtold 1994 and to the blanketing effects which
are stronger at low resolution.

The new value we have found is remarkably close to the one predicted
for the integrated contribution of QSOs, in particular once dust
obscuration of the QSO population (Meiksin \& Madau 1993; Fall \& Pei
1995) is taken into account.

Following the line of argument described in Giallongo et al. 1994, we
can use the $5\pm1$ value of $J_{22}$ together with the $\tau < 0.05$
$1\sigma$ upper limit for the GP optical depth at $z\sim 4.5$, to
derive a value for the density of the diffuse part of the
intergalactic medium, which turns out to be rather low, $\Omega_{IGM}
\mincir 0.01$.

\section{Conclusions}

A sample of absorption lines has been extracted from the spectra of 6
QSOs taken at an average resolution of 11 km s$^{-1}$ over the
absorption redshift range $z=2.8-4.1$.  Merging these data with those
of 4 other QSOs with observations of similar quality available in the
literature has provided a relatively large Lyman-$\alpha$ sample (more
than 1100 lines), to which a Maximum Likelihood analysis has been
applied to estimate simultaneously the best fit parameters of the
Lyman-$\alpha$ clouds statistics and the intensity of the UVB
background in the redshift interval $z=1.7-4.1$.

The main results of our study are:
\begin{enumerate}
\item
From the subsample of Lyman-$\alpha$ lines from the spectra at higher
S/N we derive, in the range $z=3-4$, a typical Doppler parameter in
the range $b = 20 \div 30$ km s$^{-1}$ and estimate a fraction $\sim
15-18$\% of lines with $10 \leq b\leq 20$ km s$^{-1}$, corresponding
to average temperatures $T\sim 15000$ K, still consistent with models
of low density, highly ionized clouds .
\item
We confirm the presence of a break in the column density distribution
at $\log N_{HI}=14$ with a flat slope $\beta _f\simeq 1.4$ for the
lower column densities and a steep one, $\beta _s \simeq 1.8$, for the
higher column densities.
\item
An interesting similarity is revealed between the volume density
distributions of Lyman Limit systems, Damped systems and
Lyman-$\alpha$ clouds, as derived on the basis of their estimated
sizes and geometries.  This suggests a physical association between
the Lyman-$\alpha$ clouds with $\log N_{HI} \magcir 14$ and the halos
of protogalactic systems.
\item
The intensity of the UV background $J$ as derived from the proximity
effect is a factor of 6 lower than previously estimated, with the same
photoionization model, from data at lower resolution, for which the
blanketing effects are stronger.  The value $J_{-22}\simeq 5\pm 1$ we
have found is consistent with the one predicted for QSOs after
allowing for some dust obscuration of the QSO population.
\item
No evidence is found, within relatively large uncertainties, for
evolution in redshift of the UVB in the range $z=1.7-4.1$.
\end{enumerate}

\acknowledgments
We thank K. Lanzetta, L. Lu and P. Madau for useful discussions.

\bigskip\bigskip

\newpage
\begin{center}

Table 1\\
QSO Spectra from the ESO KP\\
\begin{tabular}{cccccc}
\ \\
\hline
\hline
QSO Name & $z_{min}$ & $ z_{em}$ & FWHM (km s$^{-1}$) & Mag \\
\hline
\ \\
$2126-15$  & 2.51 & 3.27 & 11 & V=17.3 \\
$2355+01$  & 3.12 & 3.39 & 9 & V=17.5 \\
$0055-26$  & 2.96 & 3.67 & 14 & V=17.5 \\
$1208+10$  & 3.57 & 3.82 & 9 & V=17.5 \\
$1108-07$  & 3.72 & 3.95 & 9 & R=18.1 \\
$0000-26$  & 3.60 & 4.12 & 12 & V=17.5
\end{tabular}

\bigskip\bigskip

Other QSO Spectra\\
\begin{tabular}{ccccc}
\ \\
\hline
\hline
QSO Name & $z_{min}$ & $z_{em}$ & FWHM (km s$^{-1}$) & Mag \\
\hline
\ \\
$1331+17^a$ & 1.68 & 2.10 & 18 & V=16.9\\
$1101-26^b$ & 1.84 & 2.15 & 9 & V=16.0\\
$2206-19^c$ & 2.09 & 2.56 & 6 & V=17.3\\
$0014+81^d$ & 2.70 & 3.41 & 23 & V=16.5\\
\end{tabular}
\end{center}
References:
$^a$ Kulkarni et al. 1995,
$^b$ Carswell et al. 1991,
$^c$ Rauch et al. 1993,
$^d$~Rauch et al. 1992 

\begin{center}
\bigskip
\bigskip
\newpage

Table 2\\
Maximum likelihood analysis for lines with $\log N_{HI}\geq 13.3$\\
\begin{tabular}{cccccc}
\ \\
\hline
\hline
N$_l$& $\gamma$ & $\beta_{f}$& $\log N_{\infty,b}$ & $\beta_{s}$ & $J$\\
\hline
\ \\
1128& 2.49$\pm$0.21 & 1.10$\pm$0.07& 14.00$\pm$0.02& 1.80$\pm$0.03
& -21.21$\pm$0.07\\
\end{tabular}

\bigskip
\bigskip

Table 3\\
Maximum likelihood analysis for lines with $\log N_{HI}\geq 13.3$\\
with blanketing corrections
\begin{tabular}{cccccc}
\ \\
\hline
\hline
$\gamma$ & $\beta_{f}$& $\log N_{\infty,b}$ & $\beta_{s}$ & $J$
& $j$\\
\hline
\ \\
2.65$\pm$0.21 & 1.35$\pm$0.07& 13.98$\pm$0.04& 1.80$\pm$0.03 
& -21.32$\pm$0.08& --\\
2.67$\pm$0.26 & 1.34$\pm$0.09& 13.96$\pm$0.07& 1.80$\pm$0.04 
& -21.32$\pm$0.10 & -0.28$\pm$1.41\\
\end{tabular}
\end{center}

\newpage
\centerline{FIGURE CAPTIONS}

\bigskip
\noindent
Fig. 1. Examples of fits in the Lyman-$\alpha$ forest together with the 
corresponding Lyman-$\beta$ forest.

\bigskip
\noindent
Fig. 2. Normalized spectra of the regions near the Lyman-$\alpha$ emissions
for PKS 2126-158 (upper plots) and Q0000-26 (lower plots).
The dashed line shows the noise level per resolution element.

\bigskip
\noindent
Fig. 3. Doppler parameter - column density plane for the lines
out of 8 Mpc from the QSO 0000-26.

\bigskip
\noindent
Fig. 4. Doppler parameter - column density plane for the lines
out of 8 Mpc from the QSO 2126-158.

\bigskip
\noindent
Fig. 5. Doppler distributions of the lines shown in Figs. 3 and 4(continuous 
histogram for 0000-26).

\bigskip
\noindent
Fig. 6. Column density distribution for the overall composite sample
(continuous histogram). The dashed part shows the blanketing correction.

\bigskip
\noindent
Fig. 7. a,c) Blanketing simulation for line pairs. b,d) Fraction of
acceptable fits ($P>0.05$) with a single line as a function of the
separation of the line pair.

\bigskip
\noindent
Fig. 8. Redshift distribution of the Lyman-$\alpha$ lines with $\log
N_{HI}\geq 14$. At $z<1.7$ the number density of the low resolution
HST sample is shown for lines with equivalent width $W\geq 0.28$ {\AA}
The minimum value corresponds to $\log N_{HI}\simeq 14$ for $b=30$ km
s$^{-1}$ as adopted in our high resolution sample.  The redshift
distribution of the damped systems is also shown by the thick points
and that of the Lyman Limit systems by the dashed points.

\bigskip
\noindent
Fig. 9. Comoving volume density of the same samples as in Fig. 8.

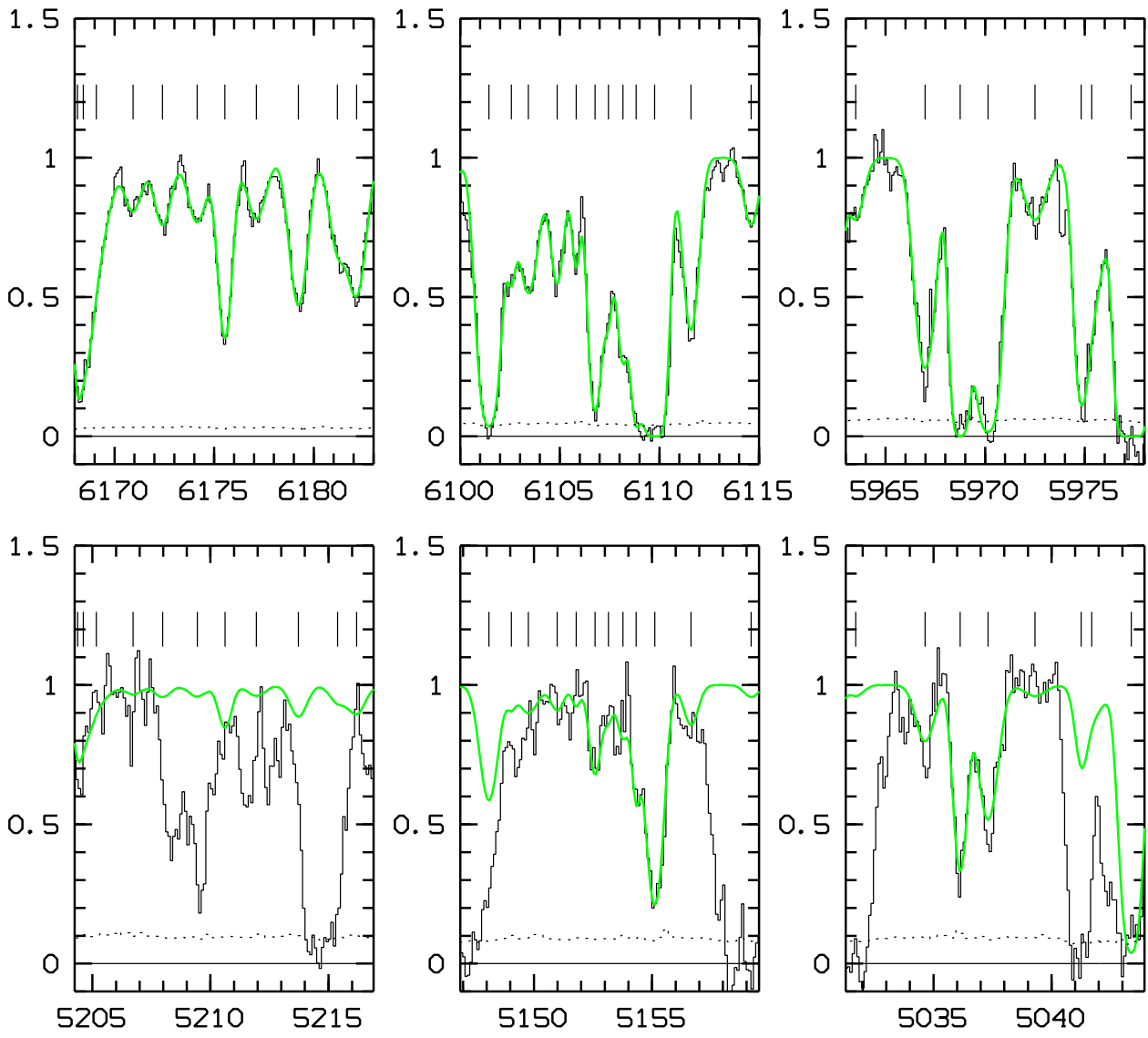
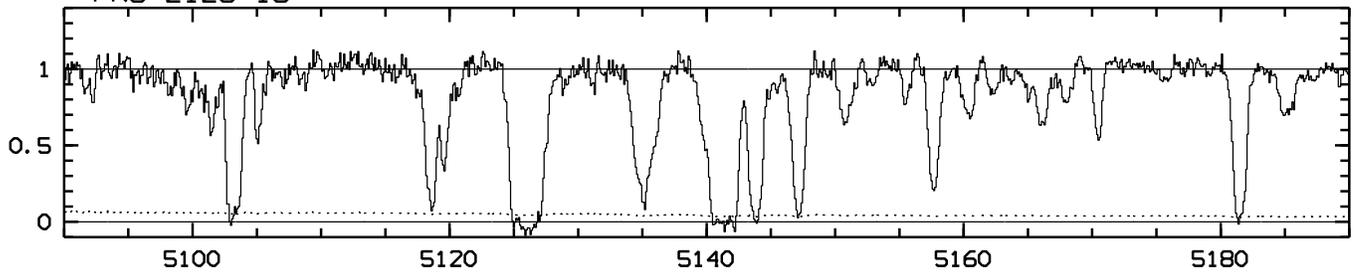
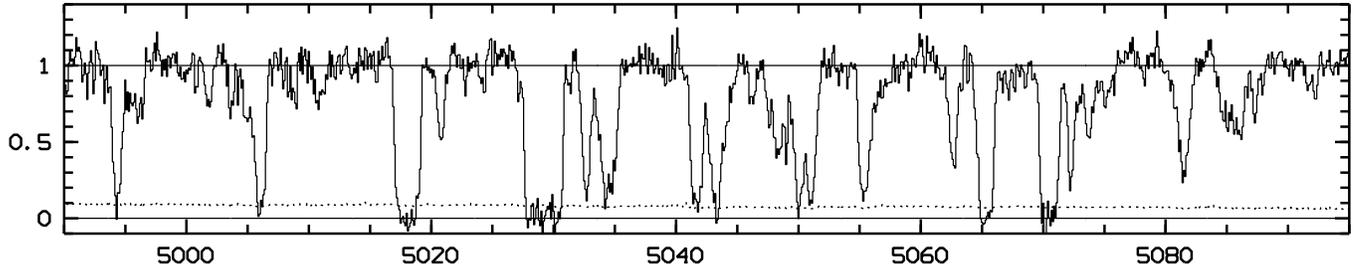
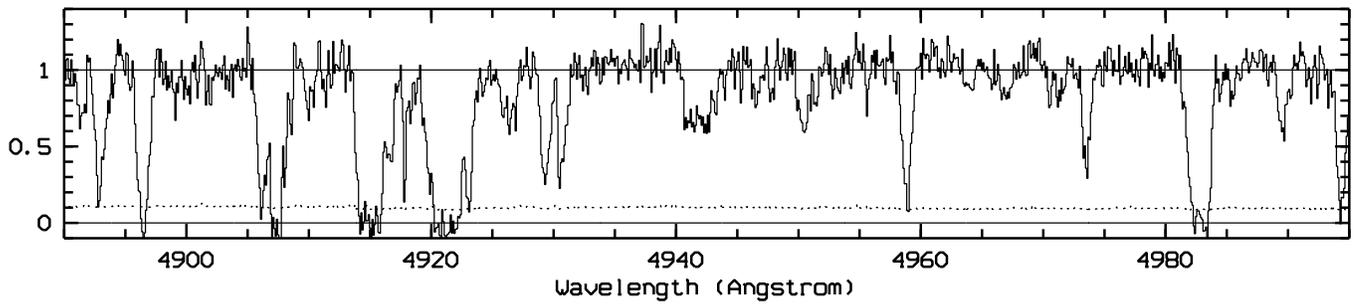
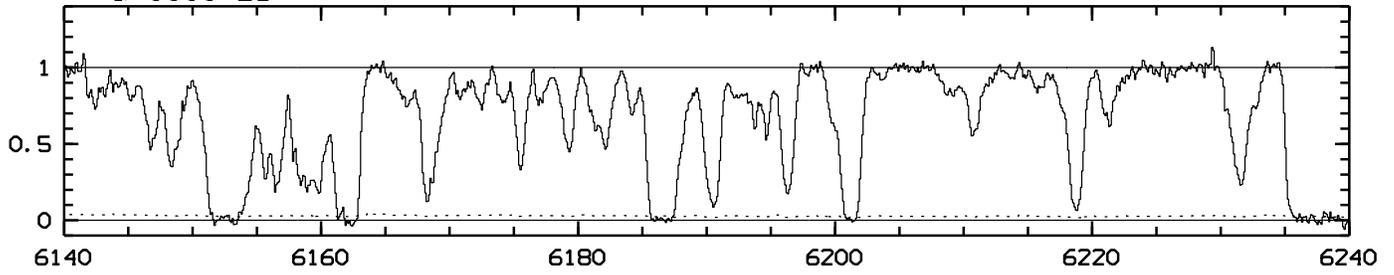
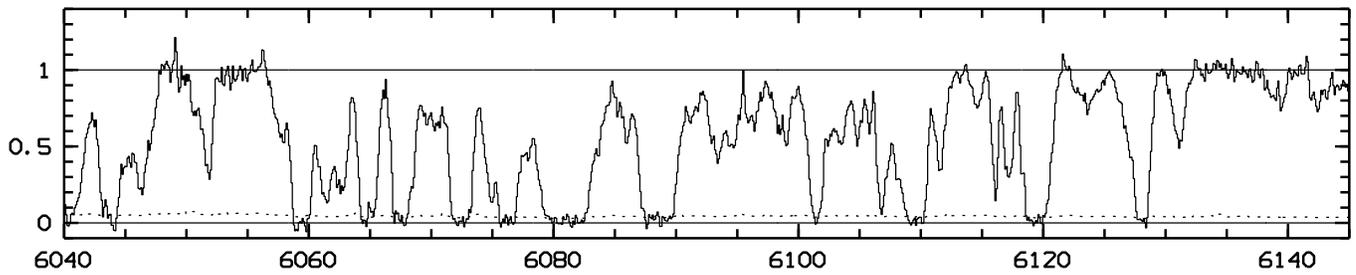
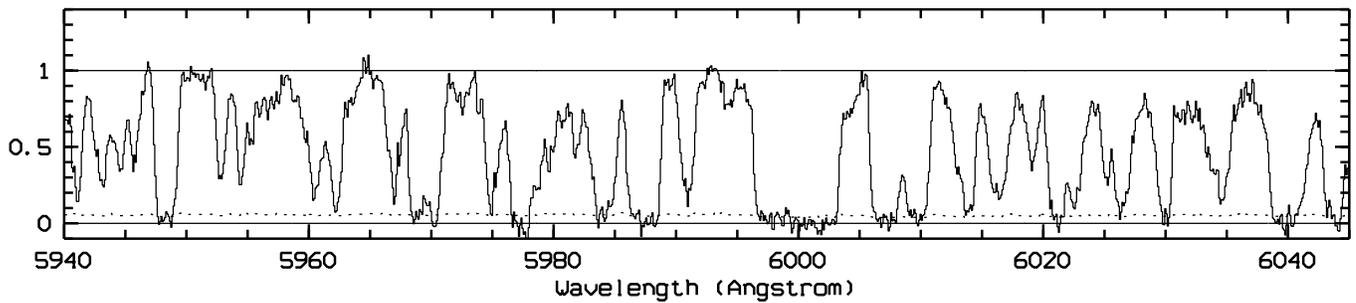
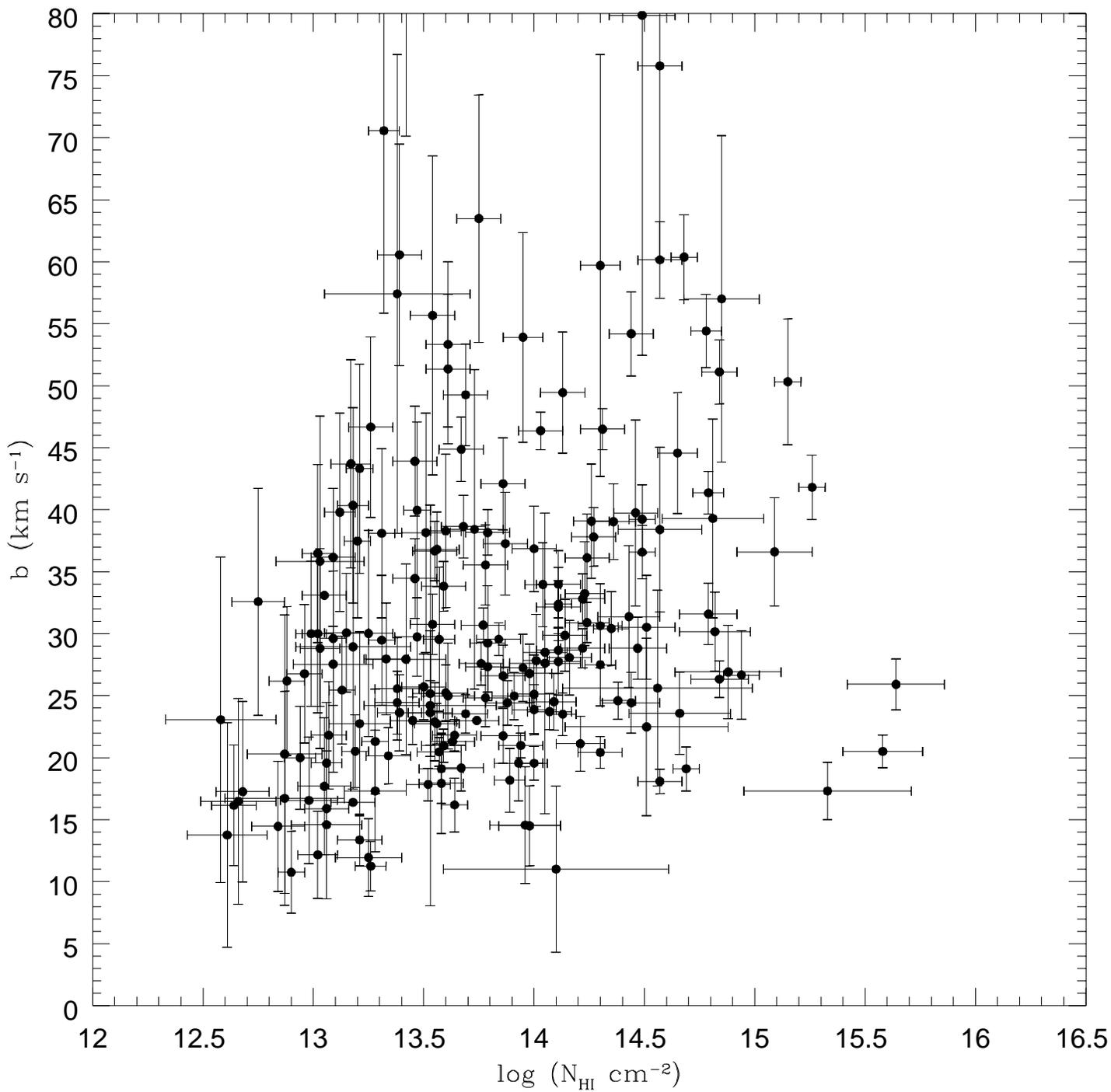
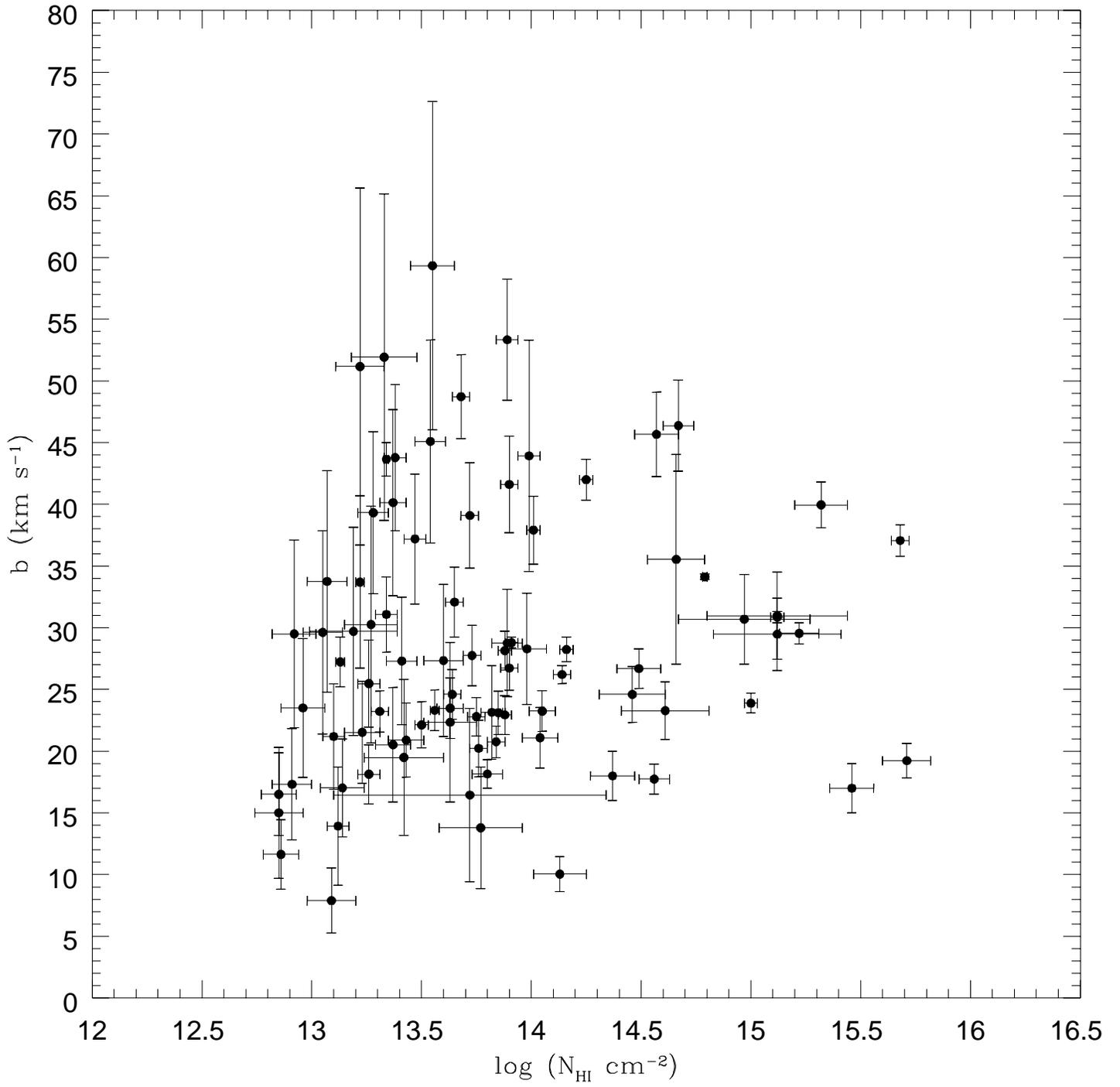
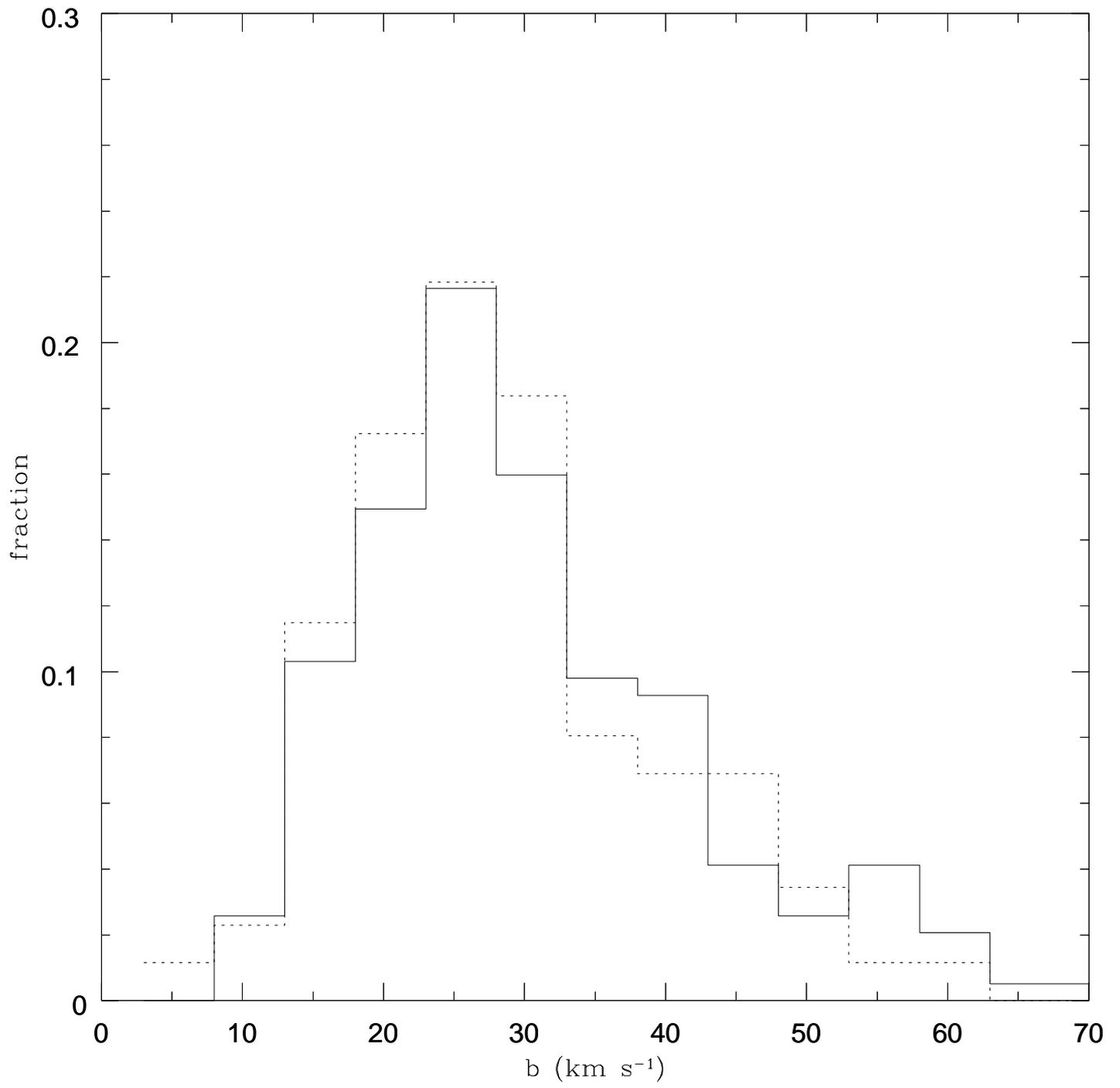
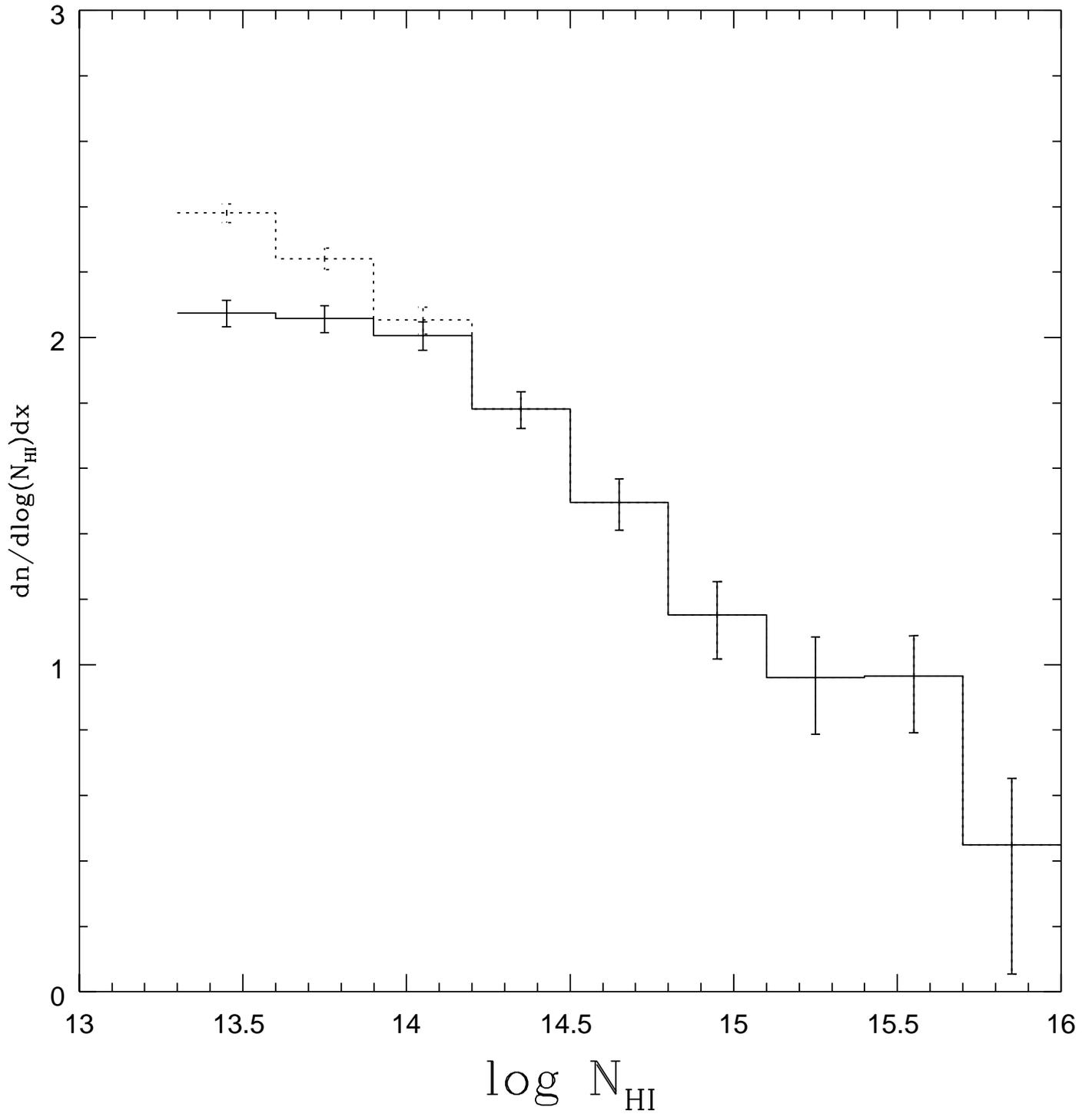
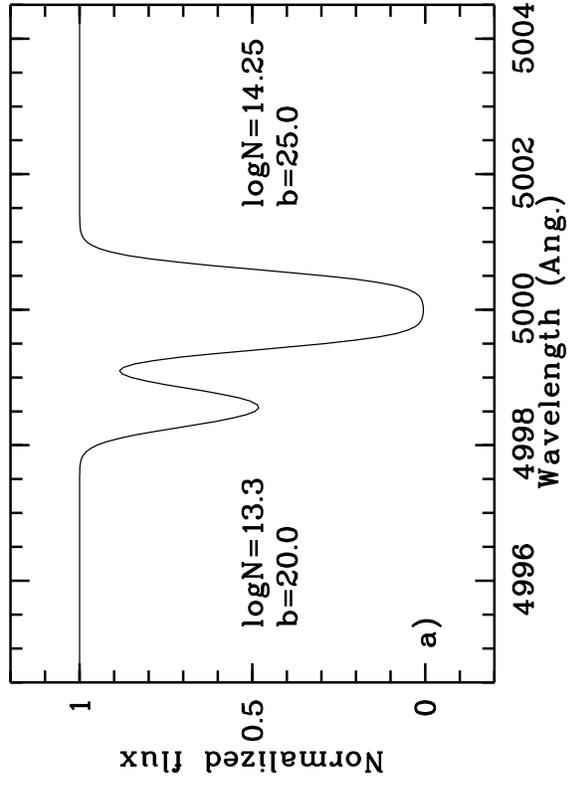
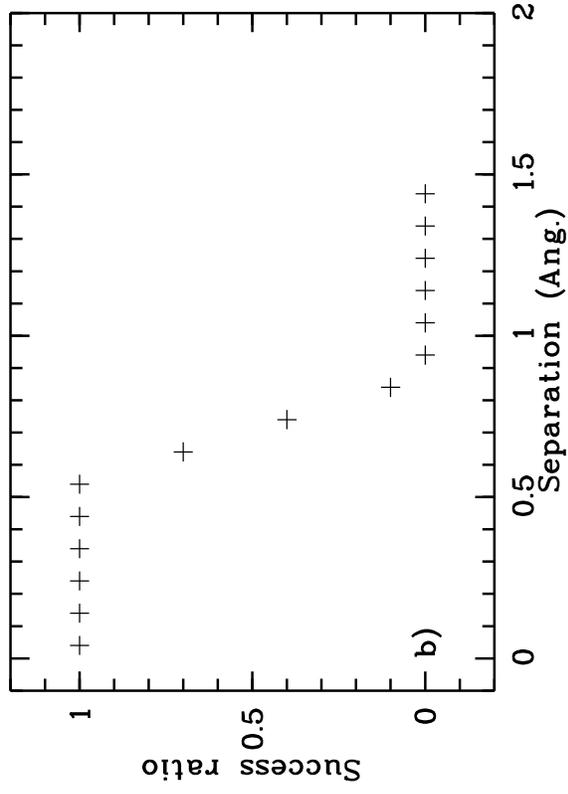
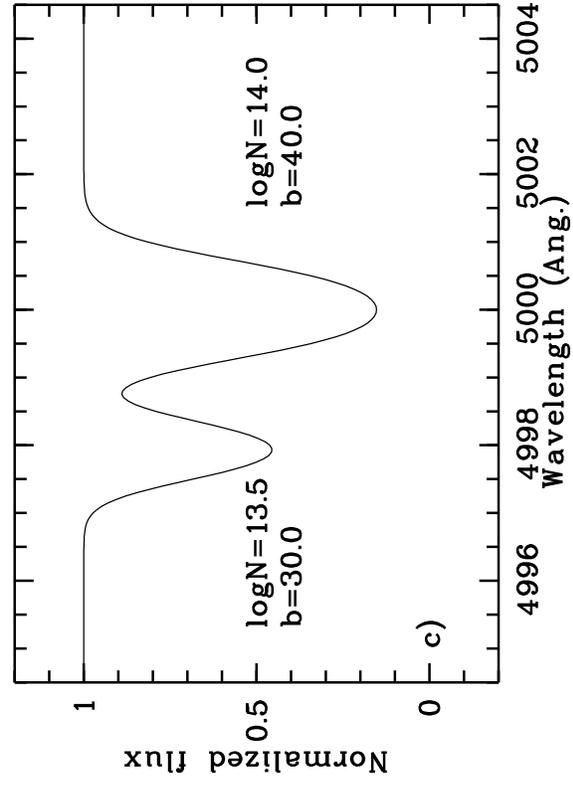
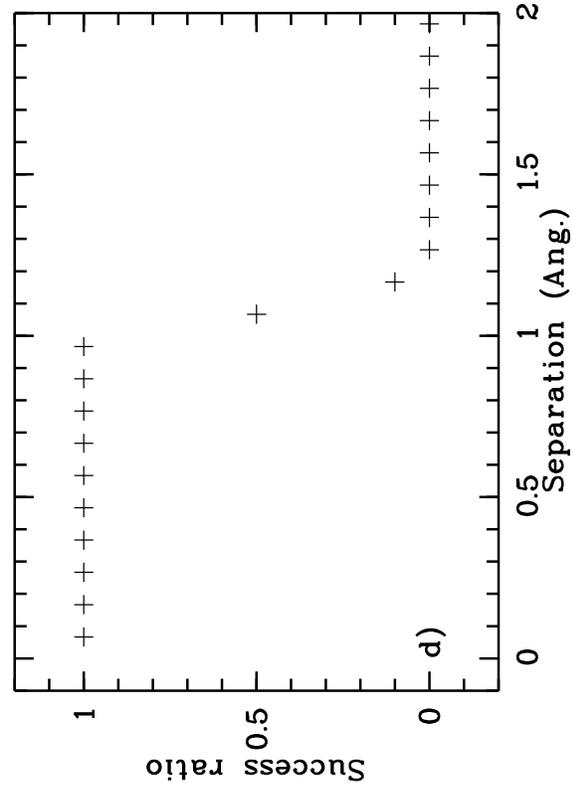
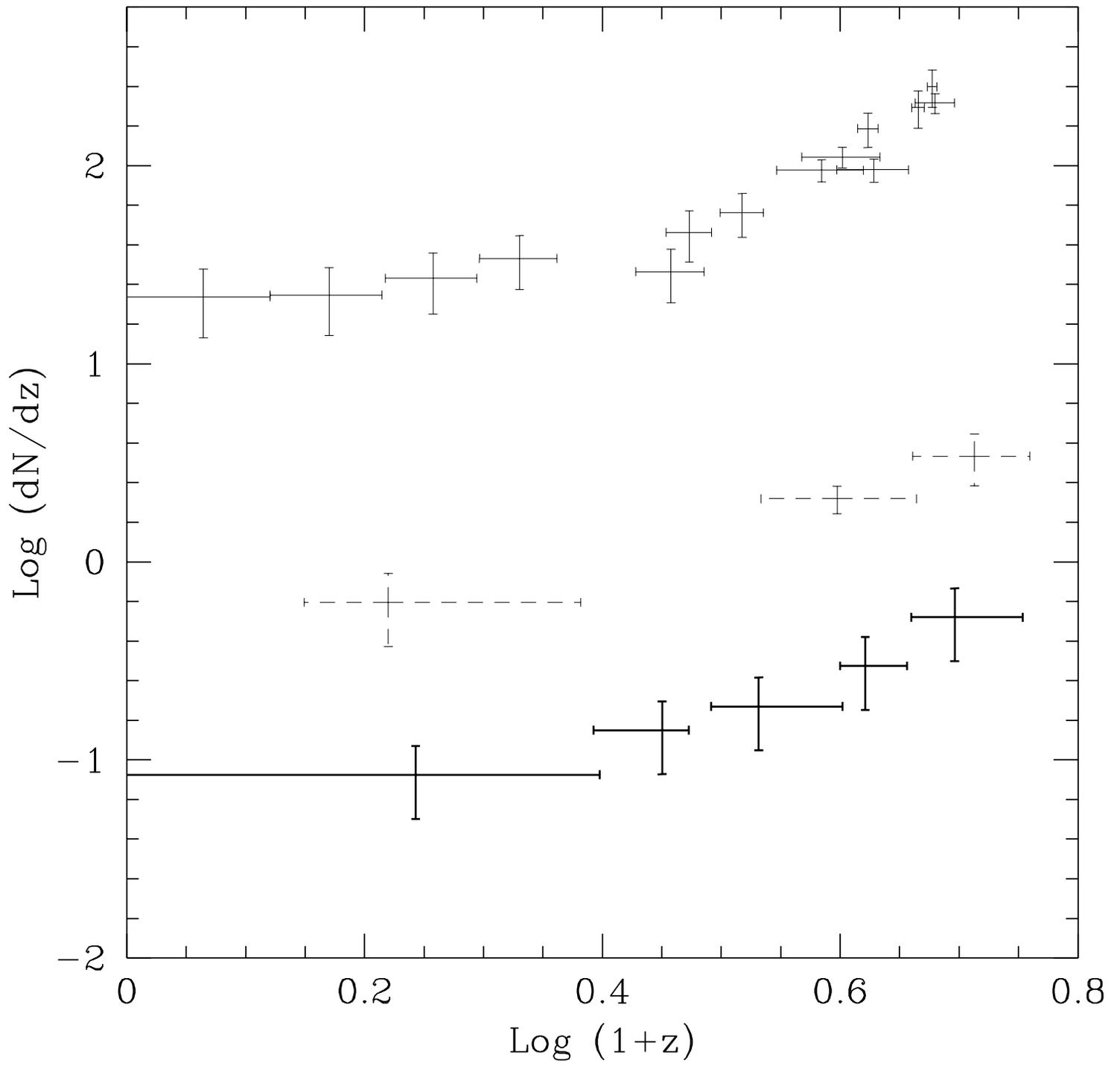
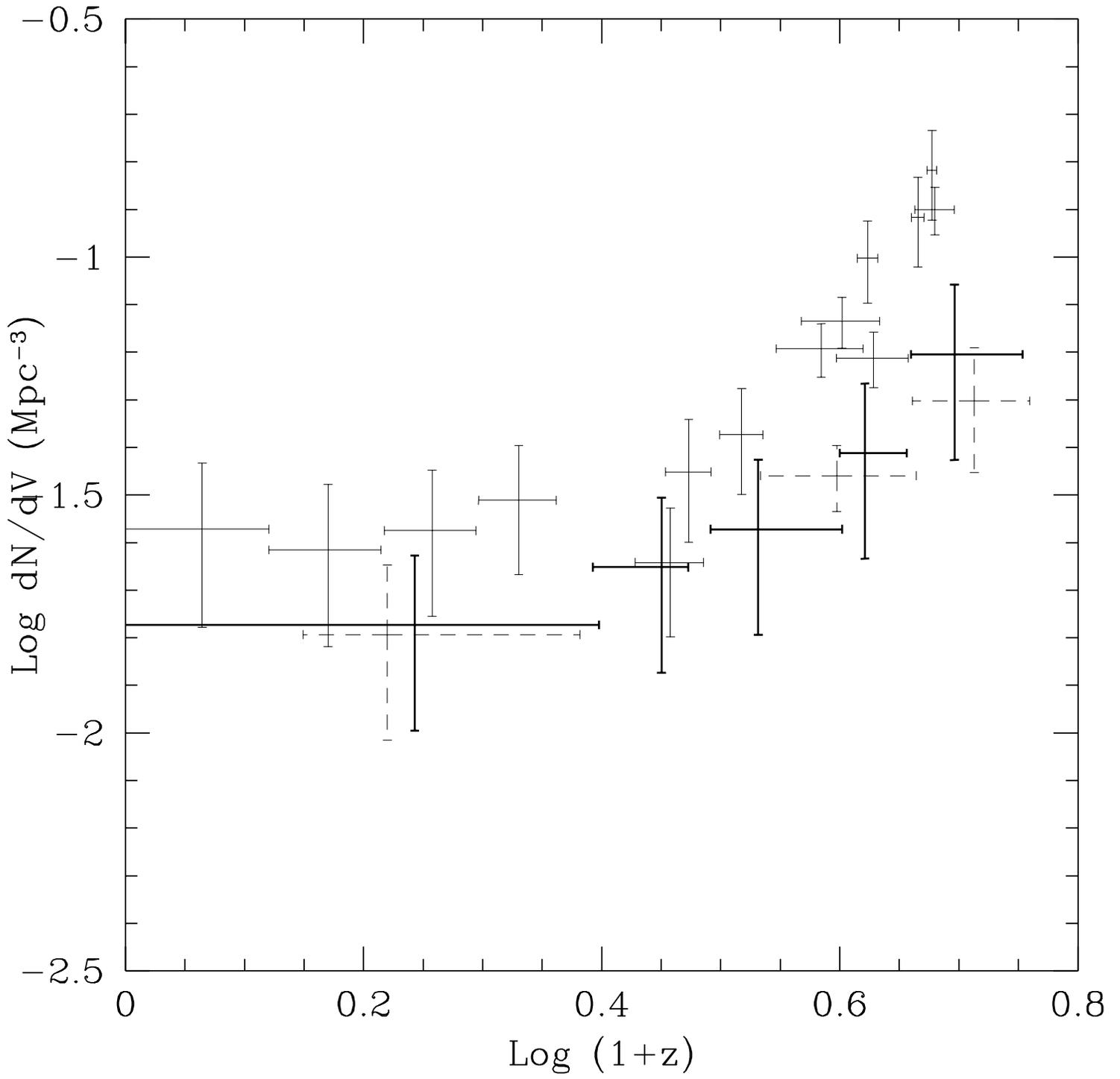
\end{document}